# Optical response of metal nanoparticle chains

Kin Hung Fung and C. T. Chan

*Department of Physics,*
*The Hong Kong University of Science and Technology,*
*Clear Water Bay, Kowloon, Hong Kong, China*
(Dated: May 3, 2006)

We study the optical responses of metal nanoparticle chains. Multiple scattering calculations are used to study the extinction cross sections of silver nanosphere chains of finite length embedded in a glass matrix. The transmission and reflection coefficients of periodic 2D arrays of silver nanospheres are also calculated to understand the interaction between nanoparticle chains. The results are in agreement with recent experiments. The splitting of plasmon-resonance modes for different polarizations of the incident light are explored. Results on the effect of disorder are also presented.

PACS numbers: 78.67.Bf, 73.22.Lp, 71.45.Gm, 78.70.-g

## I. INTRODUCTION

The excitation of surface plasmon on metal nanoparticles has attracted the attention of many researchers because of many plausible applications such as surface-enhanced Raman scattering[1,2], waveguides[3] and non-linear optics[4,5]. Recently, many techniques are developed for the fabrication of metal nanoparticles in form of clusters and ordered structures.[6,7] Metal nanoparticle chains are of particular interest because they may serve as building blocks for plasmonic optical waveguides in nanoscale[3].

Previous research has focused on metal nanoparticles with diameters from tens to hundreds of nanometers. For metal nanoparticles with diameters $d \sim 10$ nm, there were studies that focused on nanoparticle dimers[8–10] and small clusters[11,12]. For chains[13] of metal nanoparticles of such a size range, experimentally and theoretically obtained optical extinction spectra were compared by Sweatlock *et. al.* recently[14] using finite integration techniques[15] to study the length dependence of the plasmon resonance frequency of a chain of Ag nanospheres (fixing $d = 10$ nm). Here, we present the results obtained by the Multiple Scattering Theory (MST), which is highly precise numerically for spherical objects. We also consider the size dependence, chain-to-chain interaction, as well as order and disorder issues. In addition, we employ simple models to understand the optical response of chains of small metal nanoparticles.

This paper is organized as follows. In Section II, we describe the methods and the system parameters. Sec. III gives the calculated results. The extinction spectra of single silver nanoparticle chains are given with the interpretation using simple models. Then, we discuss the phenomena due to the interaction of metal nanoparticle chains and the effect of order and disorder in a chain, which is followed by the discussion and conclusion in Sec. IV.

## II. METHODS AND PARAMETERS

We consider Ag nanospheres (diameter $d$) embedded in a glass matrix of refractive index $n = 1.61$. The dielectric function of Ag spheres has the form,[8,16,17]

$$\varepsilon(\omega) = \varepsilon_a - \frac{(\varepsilon_b - \varepsilon_a)\omega_p^2}{\omega(\omega + i\gamma)}, \quad (1)$$

where $\varepsilon_a = 5.45$, $\varepsilon_b = 6.18$ and $\omega_p = 1.72 \times 10^{16}$ rad/s. This dielectric function is a fitting of the literature values[18]. The collision frequency $\gamma$ for $d$ around 10 nm has the form[19]

$$\gamma = \frac{v_F}{l} + \frac{v_F}{R}, \quad (2)$$

where $v_F = 1.38 \times 10^6$ m/s is the Fermi velocity, $l = 52$ nm is the electron mean free path at room temperature and $R = d/2$. Although there is no consensus among different authors about the proportionality between $R$ and $d$,[20] the relation $R = d/2$ is supposedly safe to use.[8]

We use MST to calculate the optical response of Ag nanosphere chains. The MST formalism can be found in details elsewhere[21,22]. It basically employs vector spherical harmonics to expand the EM fields, and the solution of the Maxwell's equations can be made as accurate as we please by increasing the angular momentum ($L$) cut-off in our expansion. For single chain of a finite number of spheres, we use the MST formulation for a finite number of particles[21,23]. For two-dimensional (2D) arrays, we use the layer MST formulation for periodic systems[22]. Unless it is specified, the truncation angular momentum $L$ in the multipole expansions is 10. We note that all numerical results in this paper are calculated using local theory only. Nonlocality issues are discussed in Sec. IV.

## III. RESULTS

### A. Single metal nanoparticle chain

We first consider a chain of Ag nanospheres of finite length with diameter $d = 10$ nm and surface-to-surface spacing $\sigma = 3$ nm (center-to-center spacing = 13 nm) embedded in the glass matrix. We set an incident electromagnetic plane wave that propagates in the direction perpendicular to the chain axis. The extinction cross sections of the chain with the number of spheres $N = 1, 2, 3, \ldots, 10$ are shown in Figure 1. When there is one single sphere, an extinction peak is found at the photon energy of about 3.0 eV. This corresponds to the Fröhlich mode of the small sphere when the dielectric constant $\varepsilon$ of the sphere matches the condition[20]

$$\varepsilon = -2\varepsilon_h, \qquad (3)$$

where $\varepsilon_h$ is the dielectric constant of the surrounding medium (glass in this case). If we do a multipole expansion of the field, the extinction peak corresponds to the "$L = 1$" mode (or dipole mode) resonance. The result shows that higher order multipoles contribute very little to the extinction cross sections for such small spheres (see Fig. 2). When $N = 2$, the extinction peak for the dimer splits into two for different light polarizations. We call the excitation mode for incident light polarized parallel (perpendicular) to the chain axis the "L" ("T") mode excitation. We see that the "L" mode extinction peak shifts towards a lower frequency (or photon energy) while the "T" mode peak shifts towards a higher freqeuncy. Similar results are obtained theoretically in Ref. 8 and 14 and experimentally in Ref. 14 and 13, and also for different kinds of metal nanoparticles[6,24]. In addition to shift of the resonance frequency, a careful examination of the result also reveal higher order ($L > 1$) contributions at about 3.1 eV, but the extinction curves are basically dominated by the dipole ($L = 1$) term.

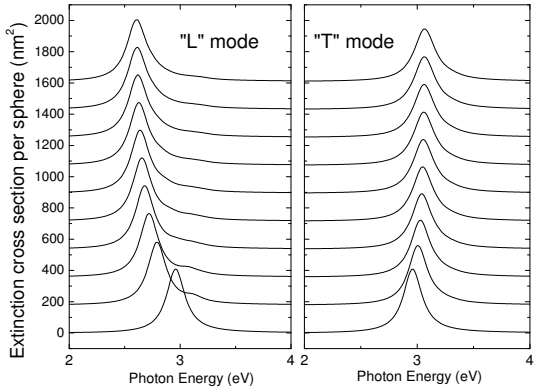

FIG. 1: Extinction spectra of a chain of Ag nanospheres with $d = 10$ nm and $\sigma = 3$ nm. Incident wavevector is perpendicular to the chain axis. The curves from bottom to top correspond to different number of spheres from 1 to 10. The left panel corresponds to "L" mode excitation (electric field parallel to the chain axis). The right panel corresponds to "T" mode excitation (electric field perpendicular to the chain axis)

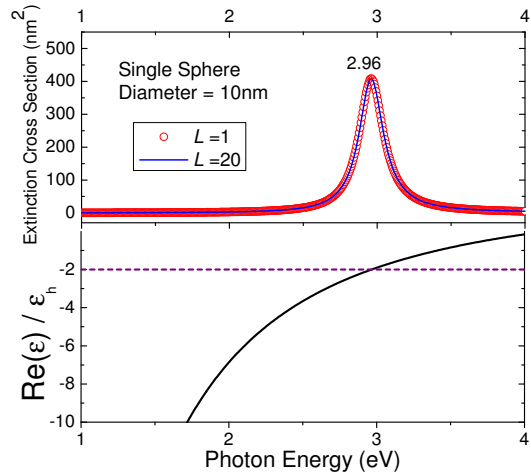

FIG. 2: (Color online) Extinction spectra of a single Ag nanosphere of diameter 10 nm. Solid line shown in the lower panel is the real part of the dielectric function divided by that of the host medium.

Here, let us first present an intuitive explanation of the shifts of the extinction peak. Fig. 3 shows a schematic induced charge distribution of a finite nanoparticle chain for both the "L" and "T" mode. For a single sphere, the natural oscillation (with frequency $\omega_0$) of the induced charges is provided by a self-induced restoring force that is due to the non-uniform surface charge distribution on its own surface, and the natural frequency is related to the surface plasmon frequency of a spherical surface. When the incident light has a frequency close to $\omega_0$ (i.e. at resonance), the extinction is significantly enhanced. It is obvious that when the spheres are sufficiently close together (but not in contact), the restoring force magnitude on the plasma in each sphere can be significantly affected by others. For "L" mode oscillation, the charge distribution on the nearest spheres attracts the charge displaced by the external field and reduces the restoring force, thus reduces the natural oscillation frequency $\omega_0$. In contrary, the charge distribution in the "T" mode oscillation increases $\omega_0$. Fig. 1 also shows that the peaks shift in the same way as we further increase $N$ to form a chain. For a short chain, a further increase in the number of spheres will further reduce (enhance) the restoring force for "L" ("T") mode (see Fig. 3). A plot of the peak frequency versus $N$ for a finite chain with $\sigma = 2$ nm, 4 nm, 6 nm is shown in Fig. 4 (a). This figure shows that, when $N$ is large, the "L" mode and "T" mode approach to an asymptotic value of about 2.5 eV and 3.09 eV for $\sigma = 2$ nm, 2.7 eV and 3.05 eV for $\sigma = 4$ nm, 2.8 eV and 3.03 eV for $\sigma = 6$ nm. Let us call these frequencies the long-chain resonance frequencies. As we can see later, these values correspond to the long wavelength collective plasmon modes of an infinite long chain of metal spheres. We found that the frequency shift of "L" mode is larger than that of "T" mode. Therefore, the "L" mode coupling is stronger than "T" mode coupling for such configuration.

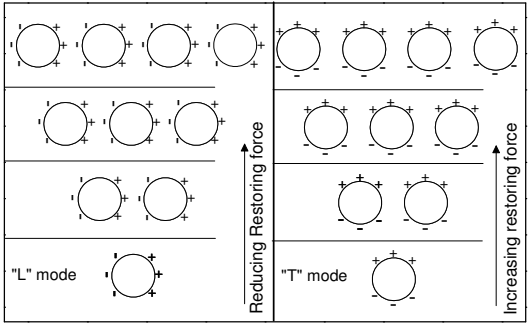

FIG. 3: Schematic diagram of the induced charge distributions on the Ag spheres at long wavelength mode. The left (right) panel corresponds to "L" ("T") mode excitation.

Corresponding figures for $d = 15$ nm and 20 nm (with the same scaling ratio in $\sigma$) are shown in Fig. 4 (b) and 4 (c). A summary of the long-chain resonance frequencies obtained here is shown in Table I. These results are qualitatively the same.

### B. Dipolar Model

The results presented in Fig. 1, 2 and 4 are (essentially exact) numerical solution of the Maxwell's equations. Here, we employ a simple model to give a semi-quantitative analysis. Since the full multiple scattering results show that the optical response is mainly dominated by the dipole ($L = 1$) terms (see Fig. 1 and 2), it should be a good approximation to employ a dipolar model. The dispersion relations of a long and infinite chain of metal nanoparticles using dipolar model has also been considered by other authors[25]. We now focus on the quasi-eigen plasmon modes of a short chain. These are actually resonance modes since the "eigen"-frequencies are complex numbers. When there is no external field, the dipole moment of the sphere $m$ induced by other dipoles is[26]

$$\mathbf{p}_m = \alpha(\omega) \sum_{m' \neq m} \frac{(1 - \beta_{m'm})\left[3\left(\hat{\mathbf{r}}_{m'm} \cdot \mathbf{p}_{m'}\right)\hat{\mathbf{r}}_{m'm} - \mathbf{p}_{m'}\right] + \beta_{m'm}^2\left[(\hat{\mathbf{r}}_{m'm} \cdot \mathbf{p}_{m'})\hat{\mathbf{r}}_{m'm} - \mathbf{p}_{m'}\right]}{r_{m'm}^3} \times e^{\beta_{m'm}}, \qquad (4)$$

where $\alpha(\omega)$ is the polarizability of the spheres, $\beta_{m'm} = \frac{in\omega r_{m'm}}{c}$, $n$ is the refractive index of the host medium (with $\mu_h = 1$), $\mathbf{r}_{m'm}$ is the position vector of sphere $m'$ measured from sphere $m$, and $c$ is the velocity of light in vacuum. For an arbitrary arrangement of non-touching spheres, Eq. 4 can be written as the general form

$$\sum_{m'} \sum_{\sigma'} M_{m'\sigma'}^{m\sigma} p_{m'\sigma'} = 0, \qquad (5)$$

where $p_{m'\sigma'}$ is the $\sigma'$-th component of the dipole moment of sphere $m'$ in Cartesian coordinates and $M_{m'\sigma'}^{m\sigma}$ is a matrix at a given angular frequency $\omega$ and arrangement of the spheres. In the case we discuss in this paper, the nanospheres are arranged as a chain so that Eq. 5 reduces to

$$\sum_{m'} M_{m'}^m p_{m'} = 0, \qquad (6)$$

where

$$M_{m'}^m = \begin{cases} \frac{1}{\alpha(\omega)} & \text{for } m = m' \\ [(1 - \beta_{m'm} + \beta_{m'm}^2) - g(3 - 3\beta_{m'm} + \beta_{m'm}^2)]\frac{e^{\beta_{m'm}}}{r_{m'm}^3} & \text{for } m \neq m' \end{cases}. \qquad (7)$$

In Eq. 7, $g = 1$ for the "L" mode and $g = 0$ for the "T" mode. In the quasistatic limit, we use $\alpha(\omega) = \frac{\varepsilon(\omega) - \varepsilon_h}{\varepsilon(\omega) + 2\varepsilon_h}\left(\frac{d}{2}\right)^3$. Here, we do not use the radiation reaction correction ($\frac{1}{\alpha(\omega)} \rightarrow \frac{1}{\alpha(\omega)} - i\frac{2\omega^3}{3c^3}$) because we are dealing with complex $\varepsilon(\omega)$.[27,28] In matrix representation, we seek

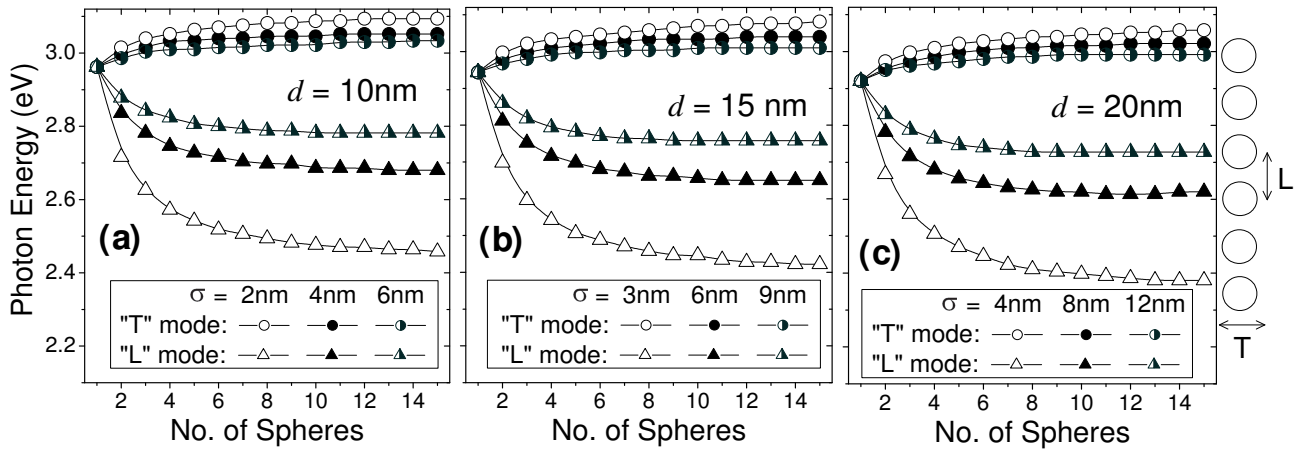

FIG. 4: Extinction peak frequency versus length of Ag nanosphere chain for different $\sigma$ and polarizations. Incident wavevector is perpendicular to the chain axis. (a) $d = 10$ nm, (b) $d = 15$ nm, (c) $d = 20$ nm.

TABLE I: Comparison among the long-chain plasmon resonance photon energies obtained in Fig. 4 for finite chain and Fig. 7 for 2D array

| $d$ (nm) | $\sigma$ (nm) | Finite chain L | Finite chain T | 2D array L | 2D array T |
|---|---|---|---|---|---|
| 10 | 2 | 2.46 eV | 3.09 eV | 2.47 eV | 3.10 eV |
| 10 | 4 | 2.68 eV | 3.05 eV | 2.69 eV | 3.07 eV |
| 10 | 6 | 2.78 eV | 3.03 eV | 2.79 eV | 3.04 eV |
| 15 | 3 | 2.42 eV | 3.08 eV | 2.45 eV | 3.10 eV |
| 15 | 6 | 2.65 eV | 3.04 eV | 2.67 eV | 3.06 eV |
| 15 | 9 | 2.76 eV | 3.01 eV | 2.77 eV | 3.02 eV |
| 20 | 4 | 2.38 eV | 3.06 eV | 2.42 eV | 3.09 eV |
| 20 | 8 | 2.62 eV | 3.02 eV | 2.64 eV | 3.05 eV |
| 20 | 12 | 2.73 eV | 2.99 eV | 2.74 eV | 3.01 eV |

for the non-trivial solutions, **p**, by solving the equation

$$\det[\mathbf{M}(\omega)] = 0. \quad (8)$$

The roots of Eq. 8 (eigen frequencies of the system), $\omega_n$, are in general complex[25] with $Im(\omega_n) < 0$. However, it may be possible to observe a resonant response at certain excitation frequency close to $Re(\omega_n)$ as long as $Im(\omega_n)$ is small. In our case, $Im(\omega_n) \approx -0.1$ eV such that $|Im(\omega_n)|/|Re(\omega_n)| < 4\%$ for all $\omega_n$ that we found.

For the 1D finite chain of Ag nanoparticles, we have compared the extinction peak frequency with $Re(\omega_n)$. Results for $N = 1$ to 7 are shown in Fig. 5 (a) and (b). When there are $N$ spheres, we get $N$ quasi-eigen frequencies satisfying Eq. 6. As the number of spheres increases and reaches the limit $N \to \infty$, the chain becomes periodic and the quasi-eigen frequencies form a band. The Brillouin zone boundary and zone center mode frequencies (for $N \to \infty$) are also shown in the figures. For the "L" mode, the band has a positive group velocity and the zone center ($k = 0$) mode has a lower frequency than the zone boundary mode. However, the opposite is true for the "T" mode band that can have a negative group velocity.[25,29] We see that the extinction peak calculated by fully-fledged MST agrees quite well with the resonance of the quasi-eigen mode at the lowest (highest) frequency for "L" ("T") mode of the dipolar model. This can be explained by the following arguments. The incident wave is a plane wave with wavevector perpendicular to the chain axis (i.e. the external incident field on all spheres are in-phase). The quasi-eigen mode with the largest number of in-phase dipoles thus contributes most resonant response because the symmetry of the mode is closest to that of the external incident field. Therefore, for "L" ("T") mode, the lowest (highest) frequency eigen mode is the one that can be excited at normal incidence (see Fig. 5 (c) and (d)).

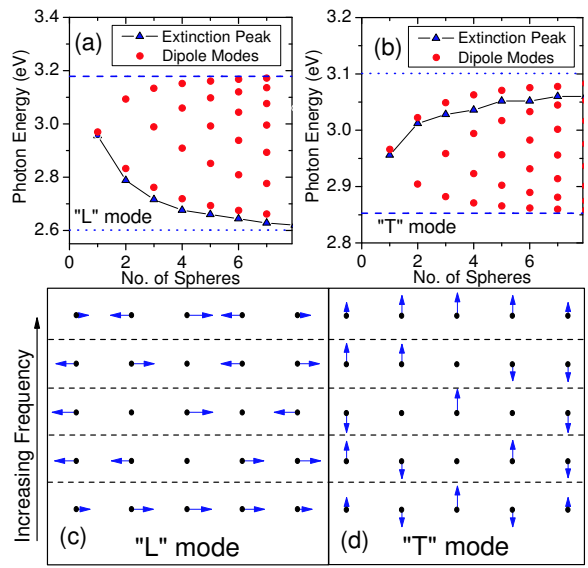

FIG. 5: (Color online) Comparison between the resonance modes (as a function of the chain length) calculated by the MST and dipolar model. Only the real part of the eigenfrequencies are shown. (a) and (b) show the quasi-eigenmode frequencies calculated by solving Eq. 8 with $d = 10$ nm and $\sigma = 3$ nm together with the extinction peak frequencies (given by MST) for "L" and "T" mode, respectively. The dashed and dotted lines are, respectively, the Brillouin zone boundary and zone center mode frequencies (for $N \to \infty$) calculated by the dipolar model. (c) and (d) show the instantaneous dipole vectors of a dipole chain with number of particles $N = 5$ at different eigenmodes.

## C. Interaction of metal nanoparticle chains

We now look at the optical response of Ag nanospheres in a two-dimensional array. The purpose of doing so is to study the interaction between nanoparticle chains. In the experiment of Ref. 13, the sample contains many coupled nanoparticle chains that are almost aligned in the same direction. Although the lengths of nanoparticle chains in the experiment may be short, it is still worth to have a quantitative study of infinitely long chains for comparison.

In this MST calculation, the sphere radii are the same as those for the nanoparticle chains we discussed. Let **a** and **b** denote the two primitive lattice vectors constructing a 2D array. For the purpose to study the chain interaction, we choose the case $\mathbf{a} \perp \mathbf{b}$. When $b >> a$, the system will act like an array of independent infinitely long nanoparticle chains. When $b \approx a$, the system can be considered as a collection of coupled nanoparticle chains. Calculated transmission, reflection and absorption spectra at normal incidence for $a = b$ are shown in Fig. 6. We see that the dominant reflection peak, absorption peak and the transmission dip are located at the same frequency (∼2.6 eV), which corresponds to the plasmon resonant frequency. Again, the dipole response dominates in this case. There are only some small features at 3.2 eV that are due to higher order resonances. In this case, the consequence of the coupling between nanoparticle chains is a shift in the plasmon resonant frequency. By varying the ratio between **a** and **b**, we can study such frequency shift due to the coupling between nanoparticle chains in greater details. We vary the magnitude of **b** and repeat the calculations. A plot of the resonant frequencies versus $b/a$ (fixing $a$ and varying $b$) for two different light polarizations is shown in Fig. 7 (a). Let us define $\sigma = a - d$. The figure shows that as **b** increases, the "L" mode ($\mathbf{E}//\mathbf{a}$) and "T" mode ($\mathbf{E}//\mathbf{b}$) resonant peak shifts from 2.51 eV to the asymptotic values of 2.47 eV and 3.10 eV for $\sigma = 2$ nm, from 2.71 eV to 2.69 eV and 3.07 eV for $\sigma = 4$ nm, from 2.80 eV to 2.79 eV and 3.04 eV for $\sigma = 6$ nm. Corresponding figures for $d = 15$ nm and 20 nm (with the same scaling ratio in $\sigma$) are shown in Fig. 7 (b) and 7 (c). All these limits obtained from the graphs agree with the long-chain resonance frequencies obtained for the single finite chain that we have discussed in Sec. III A. This consistency is expected because the limit $b >> a$ for 2D case and the limit $N \to \infty$ for "single chain" case are both representing non-interacting nanoparticle chain(s).

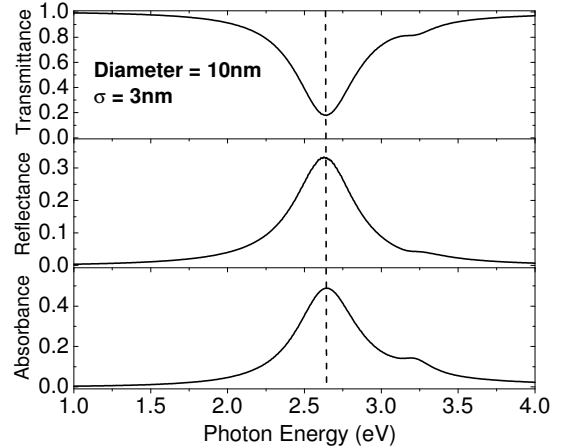

FIG. 6: Transmission, reflection and absorption spectra of Ag nanospheres arranged in square lattice $a = b = 13$ nm with $d = 10$ nm at normal incidence.

Here, we see that the splitting of resonant frequency is significant enough to be observed as long as $b/a$ is greater than about 2. This is an impoartant criteria of observing the chain properties in the experiment of Ref. 13. It is also observed that when the chains are close to each other ($b \approx a$), i.e. the spheres are arranged in square lattice, the resonant peaks are still lower than the single sphere resonance (∼2.96 eV). In this special case, assigning the "L" mode and "T" mode is in fact not suitable because both "L" mode and "T" mode couplings among nanospheres are present simultaneously. The situation then can be considered as a competition between "L" and "T" mode. From the fact that "L" ("T") mode tends to reduce (increase) the resonance frequency, we see that "L" mode coupling wins over "T" mode coupling. This is also consistent with Sec. III A that "T" mode coupling is much weaker than "L" mode coupling. Therefore, it suggests that the extinction peak will be lower in general when the nanospheres are close but not in contact ($\sigma \sim 3$ nm). Indeed, we found that this is true for many other 2D structures that we have considered, but this will not be discussed in details in this paper because we now focus on the chain properties. Another interesting feature is that the curves for "T" mode in Fig. 7 cross the point at $b/a \sim 1.8$ with photon energy 2.98 eV. This frequency is very close to the single sphere resonant frequency. This is where the competition between "L" and "T" mode coupling ends in a draw.

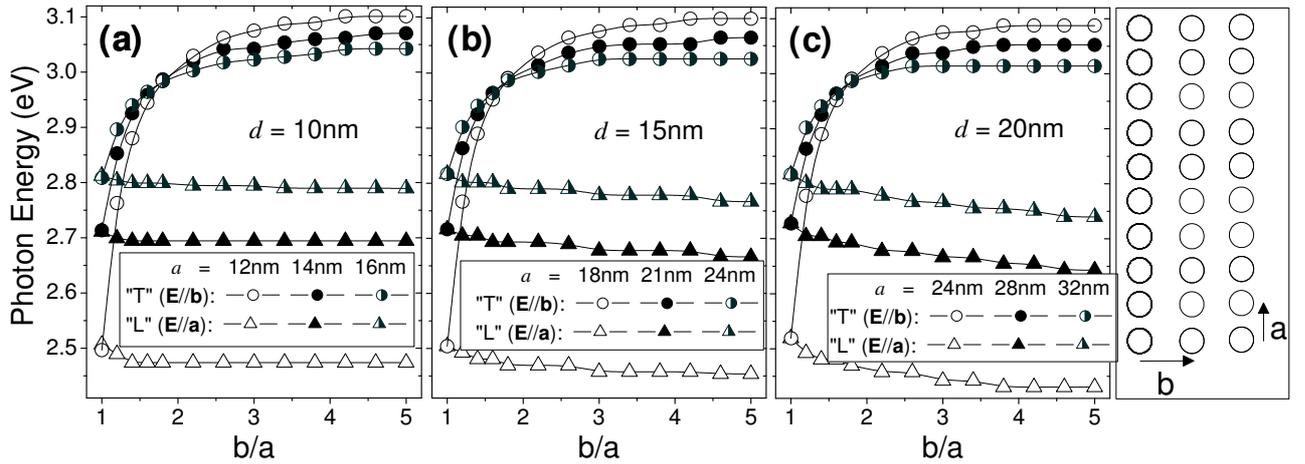

FIG. 7: Resonance frequency obtained from the reflection curves of a 2D rectangular array of Ag nanospheres at normal incidence. **a** and **b** are the two primitive lattice vector. (a) $d = 10$ nm, (b) $d = 15$ nm, (c) $d = 20$ nm.

### D. The effect of periodicity

The Ag nanoparticle chains discussed in the previous sections are arranged such that the center-to-center distances of adjacent spheres in a chain are the same. One may be interested in the question that whether this structural order can produce an extra structural resonance other than the resonance of individual spheres. For the parameters that we used previously, the possible extra resonance should be at a very high frequency such that the wavelength of the incident light is comparable to the sphere-to-sphere distances. However, such a high frequency is out of the range of interest of the present article. Searching for such effect, we increase the center-to-center distances to 400 nm.

The extinction spectrum for a 10-sphere chain with $d$ increases from 10 nm to 100 nm in steps of 10 nm is shown in Fig. 8. We note that the dielectric functions of spheres with different diameters follow Eq. 1 and 2. When $d = 10$ nm, the extinction has no difference from that of a single sphere (except the overall amplitude of the profile). As we increase the sphere diameter, we see a gradual red shift of the "$L = 1$" (Fröhlich mode) resonance peak. This is a general feature of the Fröhlich resonance.[20] Higher order ($L > 1$) resonances also appear near 3 eV when $d \geq 40$ nm, and can be traced to the higher order resonances of a single sphere. Apart from these known effects, we can see an additional feature that is due to the structural order. When we increase the sphere diameters so that $d > 80$ nm, an extra sharp peak appears at about 1.88 eV for "T" mode excitation but not for the "L" mode. Such photon energy corresponds to a wavelength of 410 nm (close to the sphere-to-sphere distance) in the host material. This feature[30–32] is due to the constructive interference of the scattered waves from the nanospheres and cannot be seen from the extinction spectra of a single sphere. For convenience, we call this peak a Bragg resonant peak.

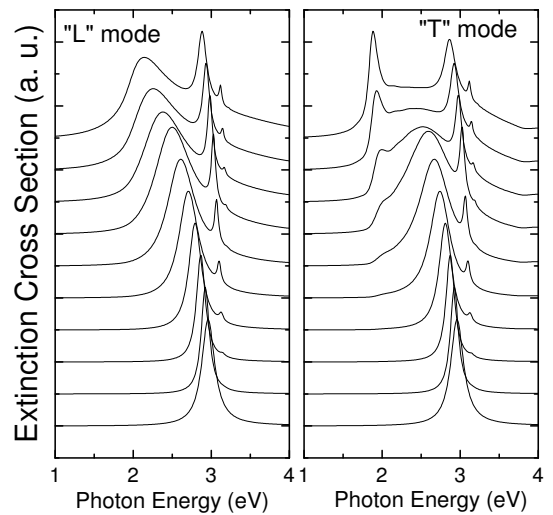

FIG. 8: Extinction spectra of Ag nanospheres chain consisting of 10 spheres with center-to-center separation $a = 400$ nm. Incident wavevector is perpendicular to the chain axis. The curves from bottom to top correspond to different diameters from 10 nm to 100 nm in 10 nm increment. Each curve is normalized in different scale so that the resonance frequencies can be easily compared.

Here, we have two points to note. The first one is why "small" Ag nanospheres cannot make a Bragg peak. The reason is that the extinction spectra of "small" Ag nanospheres are dominated by absorption while that of large Ag nanospheres are dominated by scattering.[6] The Bragg peak that we are searching is a consequence of strong scattering. Therefore, this Bragg peak does not appear for $d$ around 10 nm. Secondly, the strength of the far field scattered wave of a radiating dipole decays as $r^{-2}$ in the polar direction while it decays as $r^{-1}$ in a direction perpendicular to the polar direction. This explains why only "T" mode but not "L" mode excitation can make the Bragg peak.

Using dipolar approximations, some authors shows that the Bragg peak can be extremely sharp[31,32]. However, the sharpness of such peak requires an exact periodic arrangement of nanospheres. To understand how such peak is sensitive to disorder, we consider a 20-sphere linear array. In this example, we have an ensemble of 25 randomly generated configurations with the diameter of each sphere $d = 100$ nm and the center-to-center separation given by the expression $a = 400$ nm $+ r_1 \times 40$ nm, where $r_1$ is a random number generated uniformly from $-1$ to $1$. The particular extinction spectrum of each randomly generated nanosphere array are shown in Fig. 9. In the same figure, the averaged extinction spectra (of the 25 configurations) is also compared with that of the regular array (with $r_1 = 0$). It is clear that the resonant peaks at higher frequencies do not depend on disorder. The curves shown in the figure are basically the same in this frequency range (2.25 eV to 3.5 eV). It is because these peaks are the higher order resonances of individual sphere. In contrary, the Bragg peak is significantly affected by the orderliness of the chain (see Fig. 9 (upper panel)). While the Bragg peak for a perfectly ordered chain can be much sharper than the higher order plasmon resonance peaks, this is only true when we have perfect chains. Disorder will smooth out the sharp Bragg features (see Fig. 9 (lower panel)). We note that the Bragg peak frequency is inversely proportional to the adjacent sphere-to-sphere distance ($\omega_B \sim 1/a$). For a small change in distance $\delta a$, we have $\omega_B + \delta\omega \sim 1/(a+\delta a)$, from which we see that a positive $\delta a$ gives a smaller red shift $|\delta\omega|$ while a negative $\delta a$ of the same magnitude gives a larger blue shift $|\delta\omega|$. Therefore, in the case that $r_1$ being generated uniformly from $-1$ to $1$, the Bragg peak of the averaged curve has a blue shift.

### E. Disordered finite chain

In the previous section, we showed that the sharpness of the Bragg resonance peak is sensitive to random dislocations of the nanospheres. However, such resonance contributes very little to the extinction spectra for the values of $\sigma$ and $d$ used in Sec. III A. It thus suggests that the qualitative feature of extinction spectra should remain even if there is some randomness in the surface-to-surface distances, $\sigma$, of adjacent spheres. To see the robustness of the plasmon resonant features that we obtained, we have also studied an ensemble of 25 randomly generated configurations for each polarization. There is a chain of 10 nanospheres in each configuration. Diameter of each sphere and the surface-to-surface distance are given by, respectively, the expressions $d = 10$ nm $+ r_1 \times 2$ nm and $\sigma = 3$ nm $+ r_2 \times 1$ nm, where $r_1$ and $r_2$ are random numbers generated uniformly from $-1$ to $1$. Note again that the dielectric functions of spheres with different diameters are different according to Eq. 1 and 2. The particular and averaged extinction spectra are shown in Fig. 10. It is clear that the difference of the averaged curve from the ordered chain is very little. The profile as long as the center of peak and the peak height are almost not affected by disorder. An observable difference shown in Fig. 10 (b) for "L" mode is a red shift. This can be understood by referring back to Fig. 7. The results in Fig. 7 show that, for a fixed chain length, the red shift in the peak frequency with $\sigma$ reducing from 4 nm to 2 nm is larger than the blue shift with $\sigma$ increasing from 4 nm to 6 nm. It suggests that reducing $\sigma$ has a larger effect than increasing $\sigma$ when the amounts of changes in $\sigma$ are the same. This explains the red shift in Fig. 10 (b). Using the similar arguments, there should also be a blue shift in Fig. 10 (d) for "T" mode, but the shift is too small to be observed from the figure. Nevertheless, the little dependence on disorder shows the reason why the resonant features are still experimentally observable as in Ref. 13.

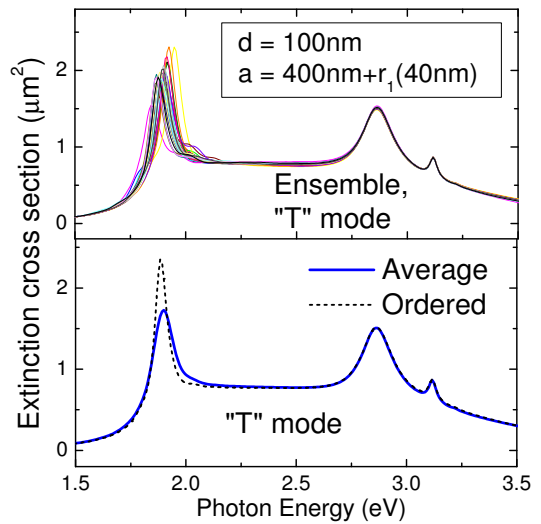

FIG. 9: (Color online) Extinction spectra of Ag nanosphere chain consisting of 20 nanospheres for "T" mode. The upper panel shows 25 particular spectra for $d = 100$ nm and $a = 400$ nm $+ r_1 \times 40$ nm, where $r_1$ is a random number generated uniformly from $-1$ to $1$. The lower panel shows the average spectrum and the spectrum corresponds to $r_1 = 0$.

## IV. DISCUSSION & CONCLUSION

Using MST, we studied the optical response of Ag nanoparticle single chains. Our results show that, for both longitudinal and transverse excitation, the shifts of the surface plasmon resonant frequency of a short Ag nanoparticle chain as we increase the chain length are monotonic and bounded. We have shown that these properties can be understood semi-quantitatively using simple models. It is also shown that the nanoparticle chain properties are already observable when the chain-to-chain spacing is greater than twice of the particle-to-particle spacing in a chain. This result and the disorder issues discussed in Sec. III E explain why the splitting phenomena of plasmon resonant band can be observed experimentally.

The numerical results presented in this paper are the full multiple scattering calculations which are based on the Mie theory. We note that we have not yet added any nonlocal correction in our calculations. Pack *et. al.* shows that nonlocal effect due to excitation of volume plasmon is negligible for two silver spheres of diameter $d = 10$ nm when the surface-to-surface spacing $\sigma > 0.4$ nm.[10] In our present work, $\sigma > 2$ nm in glass (corresponding to $\sigma > 3.2$ nm in air) for all the cases we considered. Therefore, this nonlocal effect should not be a major correction on our results. Of course, other nonlocal effects are important for nearly contact nanoparticles. This should be considered in the future.

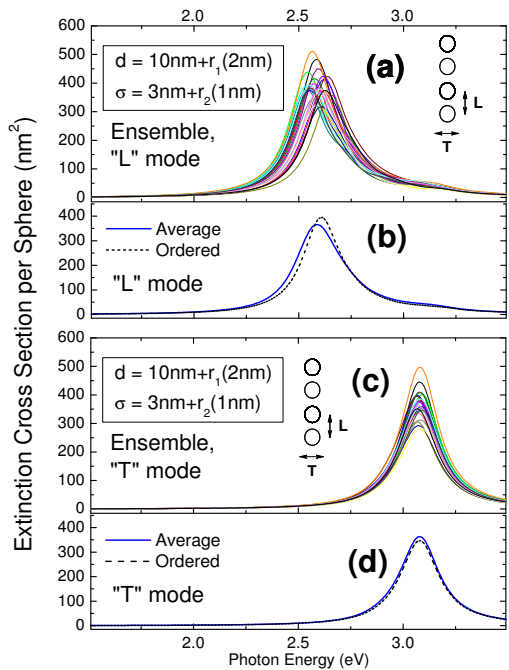

FIG. 10: (Color online) Extinction spectra of Ag nanosphere chain consisting of 10 nanospheres. Panels (a) and (c) show 25 particular spectra for "L" and "T" excitations with $d = 10$ nm $+ r_1 \times 2$ nm and $\sigma = 3$ nm $+ r_2 \times 1$ nm, where $r_1$ and $r_2$ are random numbers generated uniformly from $-1$ to $1$. Panels (b) and (d) show the averaged spectrum and the spectrum corresponds to all $r_1$ and $r_2$ equal zero for "L" and "T" excitations, respectively.

### Acknowledgments

This work is supported by the Hong Kong RGC grant 600403. We also thank Prof. Z. F. Lin and Dr. Jack Ng for discussions.